\documentclass[aps,prb,twocolumn]{revtex4}
\usepackage{amssymb}
\usepackage{graphicx}

\begin{document}
\title{The H-T and P-T phase diagram of the superconducting phase in Pd:Bi$_2$Te$_3$}
\author{Amit and Yogesh Singh}
\affiliation{Indian Institute of Science Education and Research Mohali, Sector 81, S. A. S. Nagar, Manauli PO 140306, India}
\date{\today}

\begin{abstract}
We study the magnetic field vs temperature ($H$--$T$) and pressure vs temperature ($P$--$T$) phase diagram of the $T_c \approx 5.5$~K superconducting phase in Pd$_x$Bi$_2$Te$_3$ ($x \approx 1$) using electrical resistivity versus temperature measurements at various applied magnetic fields ($H$) and magnetic susceptibility versus temperature measurements at various applied magnetic fields ($H$) and pressure ($P$).  The $H$--$T$ phase diagram has an initial upward curvature as observed in some unconventional superconductors.  The critical field extrapolated to $T = 0$~K is $H_c (0) \approx 6$--$10$~kOe.  The $T_c$ is suppressed approximately linearly with pressure at a rate $dT_c/dP \approx -0.28$~K/GPa.

\end{abstract}
\pacs{74.10.+v, 74.25.Ha, 74.25.Bt, 74.70.Ad}

\maketitle

\section{Introduction}
\label{sec:INTRO}
In the last $7$--$8$ years, topological states of matter have been a topic of intense research with several reviews being written in this short span indicating the importance of this emerging frontier \cite{Hasan-RMP-2010,Qi-RMP-2011}.   While Dirac \cite{Hasan-RMP-2010,Qi-RMP-2011} and Weyl Fermions \cite{Xu2015a, Yang2015, Xu2015b} have been discovered in real materials, Majorana Fermions remain to be discovered.  Majorana Fermions are sought after not only for their fundamental importance but also because of their possible application in topological quantum computing \cite{Nayak2008}.  There are several proposed candidate systems which might host Majorana fermion zero modes.  These include fractional quantum Hall systems \cite{MooreRead}, unconventional superconductors \cite{ReadGreen}, and heterostructures of topological insulators, semi-metals, or semiconductors with conventional superconductors \cite{FuKane2009,Sau2010,Alicea2010}.  Topological superconductors have also been predicted to be avenues to host Majorana Fermions as emergent quasi-particles \cite{Alicea2012,Elliot2014}.  It is therefore important to study in detail the superconducting properties of various candidate topological superconductors.  Thus far there are three material systems shown to host superconductivity in a topological material: Cu intercalated Bi$_2$Se$_3$ \cite{Hor2010, Kriener2011}, the topological crystalline material Sn$_{1-x}$In$_x$Te \cite{Bushmarina1986, Erickson2009}, and Pd intercalated Bi$_2$Te$_3$ \cite{Hor2011}.  While there are now several studies reporting on the superconducting properties of the two material systems Cu$_x$Bi$_2$Se$_3$ \cite{Hor2010, Kriener2011, Wray2010, Schneeloch2015} and Sn$_{1-x}$In$_x$Te \cite{Bushmarina1986, Erickson2009, Balakrishnan2013, Saghir2014}, there is only a single report on the occurrence of superconductivity in Pd$_x$Bi$_2$Te$_3$ \cite{Hor2011}.         

Herein we report synthesis, electrical transport, and magnetic properties of this novel Pd intercalated topological insulator Bi$_2$Te$_3$.   We confirm superconductivity with a critical temperature $T_c \approx 5.5$~K\@.  Using temperature ($T$) dependent resistivity ($\rho$) measurements at various magnetic fields ($H$), and temperature dependent magnetic susceptibility ($\chi$) measurements at various $H$ and pressure ($P$), we construct $H$--$T$ and $P$--$T$ phase diagrams for superconductivity in this material.  We find that the upper critical field versus temperature data show an upward curvature with decreasing temperature which has been observed for several unconventional superconductors.  Additionally, under pressure the  $T_c$ is suppressed approximately linearly at a rate $dT_c/dP \approx -0.28$~K/GPa.

\section{Experimental Details}
\label{sec:EXPT}
Single crystals and polycrystalline samples of Pd$_x$Bi$_2$Te$_3$ ($x \approx 1$) were grown using a Bridgeman technique or by conventional solid state reaction, respectively.  Stoichiometric amounts of Pd (4N), Bi (5N), and Te (5N) were ground together, pelletized and sealed in a quartz tube under partial Ar pressure.  For crystal growth the tube was placed vertically in a box furnace and heated to 900$^o$~C in 8~hrs, kept there for 10~hrs, then slowly cooled at a rate of $2^o$~C/hr to 650$^o$~C after which the furnace was switched off and allowed to cool to ambient temperature.  For polycrystalline samples, two heat treatments at 900$^o$~C for 24~hrs each with a regrind and pelletizing step in between, were used.  The XRD patterns were obtained at room temperature using a Rigaku Geigerflex diffractometer with Cu K$\alpha$ radiation.  Chemical analysis was done using energy dispersive spectroscopy (EDS) on a JEOL SEM.  Electrical transport was measured using a Quantum Design Physical Property Measurement System (PPMS).  Magnetic susceptibility $\chi(T)$ at various pressures $P \leq 1.3$~GPa were measured using a Cu-Be pressure cell with the VSM option of a Quantum Design PPMS.

\section{RESULTS}
\label{sec:RES}
\subsection{Chemical composition and Crystal Structure}
\begin{figure}[t]
\includegraphics[width=3in]{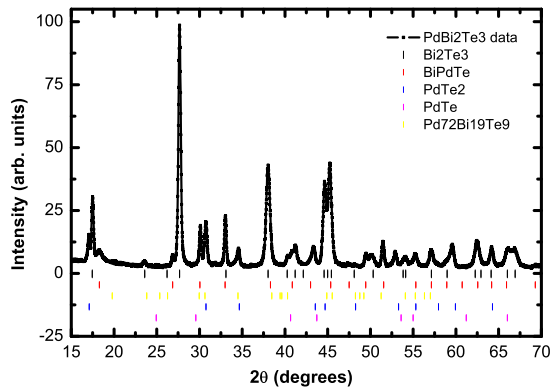}
\caption{(Color Online) Powder x-ray diffraction (PXRD) data for PdBi$_2$Te$_3$.  The vertical bars below the data are the expected Bragg positions for the various phases suspected to be present in the sample.    
\label{Fig-xrd}}
\end{figure}

Large shiny crystals could be easily cleaved from the as grown boule.  Chemical analysis gave the approximate chemical stoichimotery Pd$_{0.93}$Bi$_2$Te$_3$ which we will call PdBi$_2$Te$_3$.  Some crystals were crushed into powder for X-ray diffraction measurements. The PXRD measurements shown in Fig.~\ref{Fig-xrd} confirmed the majority phase to be Bi$_2$Te$_3$ with lattice parameters $a$~=~~4.378(5) \AA , $c$~=~30.499(7) \AA\ which are close to those of Bi$_2$Te$_3$ ($a$~=~~4.375 \AA , $c$~=~30.385 \AA ) but with a slightly enhanced $c$-axis parameter which is most likely due to Pd intercalation between layers.  We note that it is difficult to predict whether the lattice parameters will increase or decrease on intercalation, and by how much.  We point to two examples of intercalated layered materials which become superconducting.  For one material the lattice parameters increase on intercalation while for the other they decrease.  These examples are Cu$_x$TiSe$_2$ \cite{Morosan2006} and Zn$_x$ZrNCl \cite{Dompablo2000}, respectively.  For Cu$_x$TiSe$_2$ the lattice parameters increase from $a = 3.54$~\AA~ and $c = 6.01$~\AA~ to $a = 3.546$~\AA~ and $c = 6.045$~\AA~ after intercalation.  These changes are less than $0.5$~\%.  For Zn$_x$ZrNCl, the lattice parameters decrease from $a = 3.595$~\AA~ and $c = 27.640$~\AA~ to $a = 3.580$~\AA~ and $c = 27.592$~\AA~ after intercalation.  A decrease of less than $0.5$~\%.  For our samples we find an increase of similar magnitude.  The PXRD data shows the presence of secondary phases such as PdTe$_2$, BiPdTe, and
Pd$_{72}$Bi$_{19}$Te$_9$.  This is consistent with the earlier report \cite{Hor2011}.  We point out that none of the observed impurity phases are reported to superconduct above $T = 2$~K and therefore do not interfere with our goal in the present work which is to study the effect of magnetic field and pressure on superconductivity in PdBi$_2$Te$_3$.  The phase PdTe is a known superconductor with $T_c \approx 4.5$~K\@.  However, we did not observe any trace of superconductivity at $4.5$~K in our measurements.     

\begin{figure}[t]
\includegraphics[width=3in]{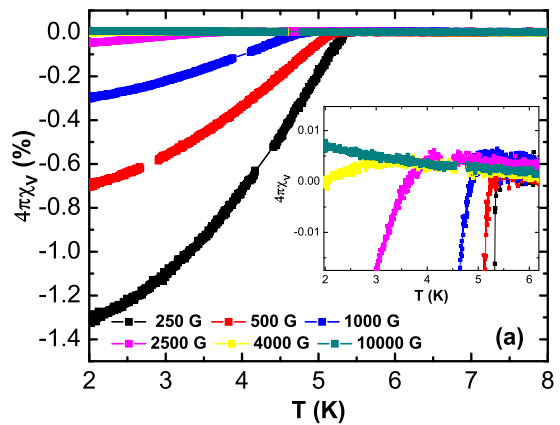}
\includegraphics[width=3in]{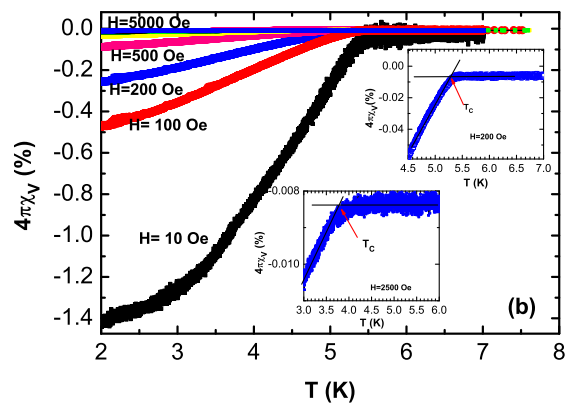}
\caption{(Color Online) Magnetic susceptibility $\chi$ versus temperature $T$ data below $T = 8$~K measured in various magnetic fields $H$ for two different samples of PdBi$_2$Te$_3$.  The data is plotted as the superconducting volume fraction $4\pi\chi_V$.  The inset in (a) shows the data on a scale so that the suppression of the onset of superconductivity with increasing magnetic fields can be seen clearly.  The inset in (b) shows the construction used to detrmine the onset $T_c$ for different $H$.  
\label{Fig-MT}}
\end{figure}

\begin{figure}[t]
\includegraphics[width=3in]{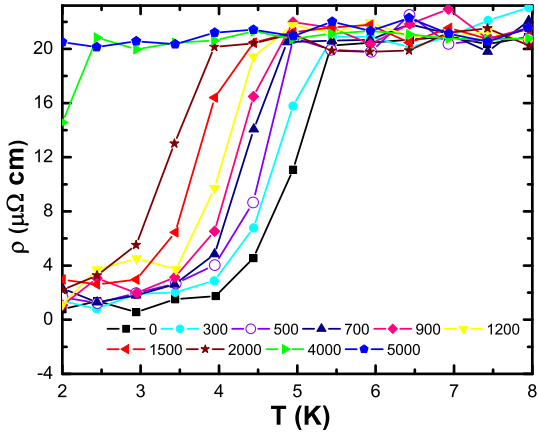}
\caption{(Color Online) Low temperature resistivity $\rho$ versus temperature $T$ for PdBi$_2$Te$_3$ at various applied magnetic fields $H$.  
\label{Fig-rho}}
\end{figure}
 
\subsection{Magnetic Susceptibility}
Magnetic susceptibility was measured for two different samples.  The percentage superconducting volume fraction (in a zero field cooled measurement) given as $4\pi$ times the volume magnetic susceptibility $\chi_V = M/H$ versus $T$ data for two PdBi$_2$Te$_3$ samples between $T = 2$~K and 8~K measured in various applied magnetic fields are shown Figs.~\ref{Fig-MT}~(a) and ~(b).  The onset $T_c = 5.5$~K and the small superconducting fraction of about $1.5$\% are consistent with the previous report \cite{Hor2010}.  The inset in Fig.~\ref{Fig-MT}~(a) shows the data on a scale where the onset of the superconducting transition for various magnetic fields can be seen clearly.  As expected, the onset $T_c$ shifts to lower temperatures with increasing magnetic fields.  Figure~\ref{Fig-MT}~(b) insets show the construction (intersection of linear extrapolations of the data above and below $T_c$) that was used to determine the onset $T_c$.  From these data, the onset $T_c$ at various magnetic fields $H$ was extracted and an $H$--$T$ phase diagram drawn.  These data are plotted in Fig.~\ref{Fig-HT}.  We will come back to a discussion of these data in a later section.  
 
\subsection{Electrical Resistivity}
 
\begin{figure}[t]
\includegraphics[width=3in]{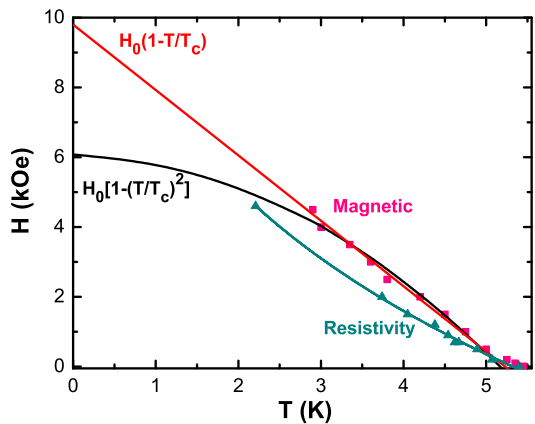}
\caption{(Color Online) The magnetic field-temperature phase diagram for PdBi$_2$Te$_3$ determined from the magnetic and electrical transport measurements.  The curves through the data are fits to different models (see text for details). 
\label{Fig-HT}}
\end{figure}

Figure~\ref{Fig-rho} shows the ac electrical resisitivity $\rho$ of PdBi$_2$Te$_3$ for $T \leq 8$~K measured at various magnetic fields.  At $H = 0$ we observe the onset of superconductivity at $T_c = 5.5$~K\@.  This onset $T_c$ is depressed to lower temperatures with increasing $H$.  The $\rho$ values do not fall to zero in the superconducting state which is consistent with the earlier report.\cite{Hor2010}.  However, the Meissner effect and the depression of $T_c$ with $H$ confirms superconductivity in PdBi$_2$Te$_3$.  The onset $T_c$ at various magnetic fields $H$ were extracted from these data and an $H$--$T$ phase diagram was drawn.  These data are also plotted in Fig.~\ref{Fig-HT}.

\subsection{Pressure Dependent Magnetic Susceptibility}
 
\begin{figure}[t]
\includegraphics[width=3in]{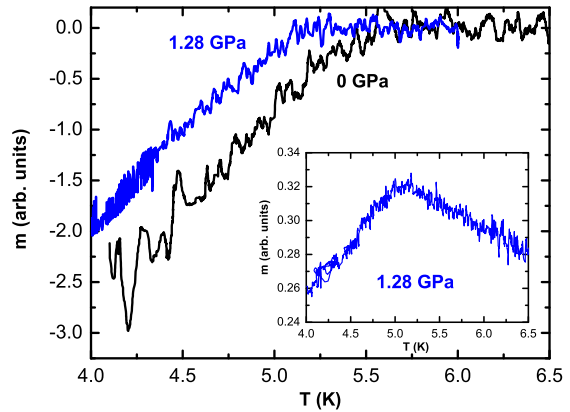}
\caption{(Color Online) Magnetization versus temperature at different pressures measured in a magnetic field of $H = 100$~Oe.  The inset shows the magnetization data of the pressure cell plus the PdBi$_2$Te$_3$ sample highlighting the abrupt downturn of the data on the onset of superconductivity.  
\label{Fig-chi-P}}
\end{figure}

\begin{figure}[t]
\includegraphics[width=3in]{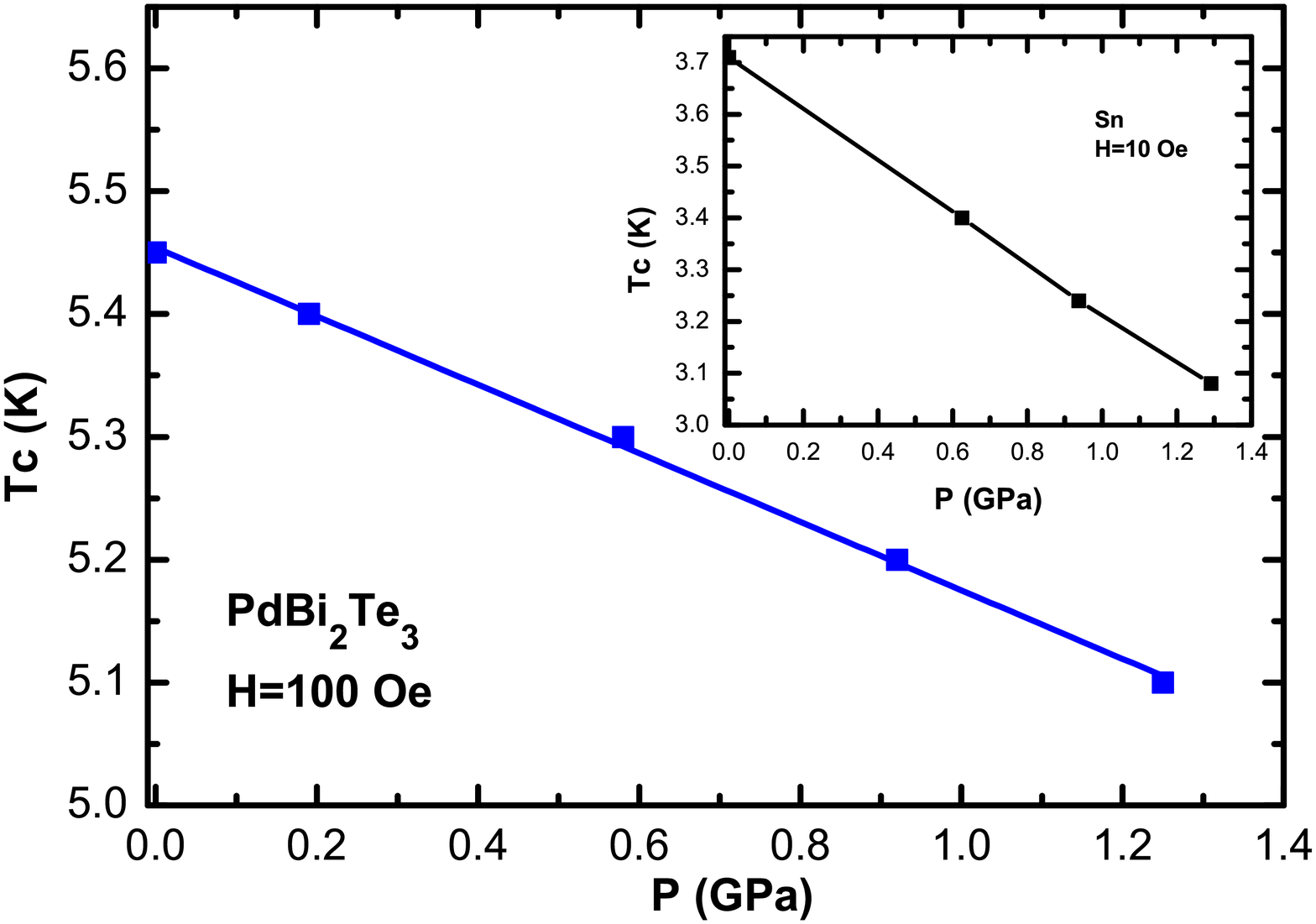}
\caption{(Color Online) Pressure $P$ dependence of the superconducting transition temperature $T_c$ for PdBi$_2$Te$_3$.  Inset shows the $P$-vs-$T_c$ for Sn which was used as a manometer.  
\label{Fig-PT}}
\end{figure}

Figure~\ref{Fig-chi-P} shows the low temperature magnetization versus temperature data at two pressures: $P = 0$ and $P = 1.28$~GPa.  In these measurements, Sn was used as a manometer. The superconducting $T_c$ for Sn (not shown) at various pressures was used to determine the pressure.  To reveal the superconductivity of PdBi$_2$Te$_3$ over the background of the pressure cell we had to apply a fairly large magnetic field of $H = 100$~Oe.  Figure~\ref{Fig-chi-P} inset shows a typical magnetization plot where the superconducting transition can be seen as a sudden downturn of the data on top of a paramagnetic signal from the cell.  This paramagnetic background signal was subtracted from the raw data to give plots shown in the main panel of Figure~\ref{Fig-chi-P}.  Figure~\ref{Fig-chi-P} shows that the $T_c$ for PdBi$_2$Te$_3$ is suppressed to lower temperatures on increasing pressure.  Similar measurements were performed at pressures $P \approx 0, 0.2, 0.6, 0.9, 1.28$, respectively.  These data were used to extract the onset $T_c$ for each $P$.  Figure~\ref{Fig-PT} shows the $P$--$T$ phase diagram for PdBi$_2$Te$_3$ for $P \leq 1.3$~GPa.  We find that the $T_c$ is suppressed approximately linearly with pressure at a rate $dT_c/dP \approx -0.28$~K/GPa.  

\section{Summary and Discussion}
In summary, we have synthesized single crystals and polycrystals of Pd intercalated Bi$_2$Te$_3$, confirmed the occurrence of superconductivity with a $T_c = 5.5$~K\@, and studied the superconducting properties under magnetic field $H$ and externally applied pressure $P$.  The magnetic field-temperature ($H$--$T$) phase diagram extracted from both electrical transport and magnetic measurements show an upward curvature near $T_c$ as has been previously observed for several unconventional superconductors like high $T_c$ cuprates and organic superconductors \cite{Zavaritsky2002, Maple1997} as well as for the multi-gap superconductor MgB$_2$ \cite{Shi2003}.  We tried to fit our $H$--$T$ data using phenomenological models that have been used previously for conventional BCS type superconductors.  Figure~\ref{Fig-HT} shows results of our attempts to fit the $H$--$T$ data obtained using the magnetic measurements described above with the expression $H(T) = H_0[1-({T\over T_c})^n]$, for fixed $n = 1$ or $2$, where $H_0$ is the $T = 0$ critical field.  It can be seen from Fig.~\ref{Fig-HT} that the fit for $n = 2$ is extremely poor and the fit for $n = 1$ is satisfactory only at lower temperatures away from the $H = 0$ critical temperature $T_c = 5.5$~K\@.  The $H_0$ obtained in these fits range approximately from $6$ to $10$~kOe.  If $n$ is allowed to vary as a fit parameter, the best fit is obtained for $n < 1$ confirming the unconventional upward curvature.  For example, the fit giving $n \approx 0.4$ is shown as the solid curve through the $H$--$T$ data obtained from the resistivity measurements.  This $H(T)$ behavior is unusual and suggests the possibility of unconventional superconductivity in PdBi$_2$Te$_3$.

Additionally, from high pressure magnetic measurements we found that $T_c$ is suppressed to lower temperatures approximately linearly with pressure $P$ at a rate $dT_c/dP \approx -0.28$~K/GPa.  This suppression of $T_c$ with pressure is expected for conventional electron-phonon mediated superconductivity and occurs due to the stiffening of the lattice with pressure.  The $H$--$T$ and $P$--$T$ phase diagrams therefore give seemingly conflicting results about the nature (conventional or unconventional) of superconductivity in PdBi$_2$Te$_3$.  We note that MgB$_2$ also shows such conflicting behaviors \cite{Shi2003, Deemyad2001}.  MgB$_2$ is an electron-phonon mediated superconductor and hence its $T_c$ decreases with pressure $P$.  The unconventional upward curvature in its $H$--$T$ phase diagram most likely arises due to its two-gap nature.  The reason for such behavior in PdBi$_2$Te$_3$ is not understood at present.

\begin{acknowledgments}
We acknowledge support of the X-ray facility at IISER Mohali for powder XRD measurements.  YS acknowledges DST, India for support through Ramanujan Grant \#SR/S2/RJN-76/2010 and through DST grant \#SB/S2/CMP-001/2013.  \\
\end{acknowledgments}

\end{document}